# Spectral sensitive phonon wipeout due to a fluctuating spin state in a $Fe^{2+}$ coordination polymer

V. Gnezdilov[1], P. Lemmens[2,*], P. Scheib[2], M. Ghosh[2], Yu. G. Pashkevich[3], H. Paulsen[4], V. Schünemann[5], J. A. Wolny[5], G. Agustí[6], J. A. Real[6]

[1]*B.I. Verkin Inst. for Low Temperature Physics and Engineering, NASU, 61103 Kharkov, Ukraine;*
[2]*Inst. for Condensed Matter Physics, TU Braunschweig, D-38106 Braunschweig, Germany;*
[3]*A.A. Galkin Donetsk Phystech NASU, 83114 Donetsk, Ukraine;*
[4]*Inst. für Physik, Univ. zu Lübeck, Ratzeburger Allee 160, 23538 Lübeck, Germany;*
[5]*Fachbereich Physik, TU Kaiserslautern, E. Schrödinger Str., D-67663 Kaiserslautern, Germany;*
[6]*Instituto de Ciencia Molecular/Departamento de Química Inorgánica Edifício de Institutos de Paterna, Apartado de correos 22085, 46071 Valencia, Spain*

## Abstract:

Raman scattering in the spin-crossover system [Fe(pmd)(H$_2$O){Au(CN)$_2$}$_2$]·H$_2$O reveals a complex three-phase spin-state transition in contrast to earlier observations in magnetization measurements. We observe different spin state phases as function of temperature and electromagnetic radiation in the visible spectral range. There exists a fluctuating spin state phase with an unexpected wipeout of the low frequency phonon scattering intensity. Furthermore we observe one phase with reduced symmetry that is attributed to a cooperative Jahn-Teller effect. Pronounced electron-phonon interaction manifests itself as a strong Fano-resonance of phonons related to {FeN$_6$} and {FeN$_4$O$_2$} coordination octahedra. Density functional theory supports this interpretation.

*: Author for correspondence.

# Introduction

Functional materials with switching properties gained considerable interest in view of their potential technological applications.[1,2] Despite numerous reports on coordination polymers with specific topologies [3], the incorporation of electronically and/or optically active building blocks as essential structural components of such functional materials has scarcely been explored.[4] In this regard, the use of spin-crossover building blocks has been shown to be a suitable strategy since they change reversibly their magnetic, structural, dielectric and optical properties in response to stimuli such as a variation of temperature or pressure and light irradiation.[5-8] In particular, six-coordinate $Fe^{2+}$ compounds, with $3d^6$ electron configuration, represent an important class of switchable molecular systems. In pseudo-octahedral symmetry, they change reversibly its ground state from an $^1A_1$ ($t_{2g}^6$), i.e. low-spin (LS) state to a $^5T_2$ ($t_{2g}^4 e_g^2$), i.e. high-spin (HS) state.

The light induced excited spin-state trapping (LIESST) effect [9,10] describes a light induced transition of the $^1A_1$ (LS) → $^5T_2$ (HS) states in $Fe^{2+}$ spin-crossover compounds at temperatures far below the thermal spin state transition temperature $T_c$. According to a two-step mechanism, postulated by *Decurtins et al.*[10], irradiation into the LS d-d bands, i.e. $^1A_1 \rightarrow ^1T_1/^1T_2$, leads to the population of the $^5T_2$ HS state via the $^3T_1/^3T_2$ state. The relation between photo-induced phase transitions and thermally induced ones, however, has only recently been highlighted. Based on the similarity of absorption spectra, magnetic properties [11], and reflection spectra [12,13] it has been argued the photo-induced phase at low temperatures is the same state as the equilibrium phase at high temperatures.

Recent experiments have shown that this assumtion has to be revised. Using resonant Raman spectroscopy, it was found by *Tayagaki et al.*[14] that at low temperatures the photo-induced phase has a broken symmetry and differs from the thermally-induced high-symmetry phase. This was illustrated more recently in the spin-crossover field by Real et al.[15] who demonstrated the occurrence of light-induced polymorphism on single crystals at low temperature. Similar conclusions have been made recently based on pico-second time-resolved crystallography [16].

According to the thermal evolution of the fraction of $Fe^{2+}$ centers in the HS state, $\gamma_{HS}(T)$, the spin change can be classified as (a) "gradual", covering a wide temperature range; (b) "stepwise" or "abrupt", with changes from one state to another within only a few Kelvin; (c) showing a hysteresis effect. The former transitions are characterized by continuous evolution of the X-ray pattern indicating that no distinguishable phases exist, whereas in the latter transitions (b) and (c) at least two independent crystallographic phases, HS and LS, can be detected. The hysteretic behavior stems from cooperative changes in the crystal and confers to these materials a memory function. Cooperativeness can be enhanced improving communication between the active spin-crossover building blocks, i.e. integrating them in rigid frameworks. For instance, the possibility to induce polymerization through suitable coordinated anions, i.e. cyanide ligands, has been considered only recently as a suitable strategy. Cyanide-bridged homo- and hetero-metallic coordination polymers have been shown to exhibit a remarkable diversity of structural types with interesting magnetic, electrochemical, magneto-optical, thermo-mechanical, and zeolitic properties.[17] In particular, Hofmann-like clathrate compounds[18] containing Fe(II) ions have opened up the opportunity to build highly cooperative thermo-, piezo-, and photo-switchable two- and three-dimensional coordination polymers [19-22].

The compound [Fe(pmd)(H$_2$O){Au(CN)$_2$}$_2$]·H$_2$O (**1Au**), pmd = pyrimidine, represents a singular coordination polymer due to its structural and electronic properties. It is made up of three identical and independent three-dimensional interlocked frameworks (Fig. 1) with two crystallographically independent {Fe(1)N$_6$} and {Fe(2)N$_4$O$_2$} sites. Both iron sites lie at inversion centers and correspond to an elongated and a compressed coordination octahedron, respectively. Interestingly, the three frameworks collapse at elevated temperatures (ca 400 K) due to a cooperative topochemical ligand substitution which changes the local coordination of the Fe(2) site to form a new 3D framework (**2Au**) [23]. Magnetic susceptibility measurements show that (**1Au**) undergoes a sharp cooperative spin-crossover with critical temperatures $T_c^{down}$ = 163 K and $T_c^{up}$ = 171 K for cooling and warming, respectively, leading to a hysteresis of $\Delta T_c$ = 8 K. Crystallographic data demonstrates that only site Fe(1) undergoes the spin-crossover. It is worth noting that the spin-crossover is accompanied by a drastic color change from pale yellow (HS) to deep red (LS). [23]

Raman spectroscopy is a well-established technique to probe spin states of spin-crossover materials [14,24,-27]. Earlier Raman spectroscopy investigations have mainly been devoted to the differentiation between high- and low-spin Fe$^{2+}$ complexes or to the estimation of the respective entropy contributions. More attention has been given recently to Raman scattering (RS) as a convenient and easily accessible probe of light induced spin transitions[28] for temperatures far below and above the spin state transition temperature. Unfortunately, much less is known to spin conversion phenomena at temperatures close to $T_c$ where a complex competition of temperature and light induced effects is expected. In addition, especially complex scenarios are possible in spin-crossover systems with more than one Fe$^{2+}$ site on nonequivalent lattice sites as given in the present coordination polymer.

Here, we report on a Raman scattering (RS) study of the two site Fe$^{2+}$ coordination polymer [Fe(pmd)(H$_2$O){Au(CN)$_2$}$_2$]·H$_2$O to establish a new phase diagram and demonstrate a more dynamic evolution of the spin state as function of temperature and light irradiation than expected from earlier magnetization data [23]. A novel phonon "wipe-out" effect is demonstrated proposing a fluctuating spin (FS) state phase in the phase diagram. Our phonon assignment is supported by density functional calculations (DFT).

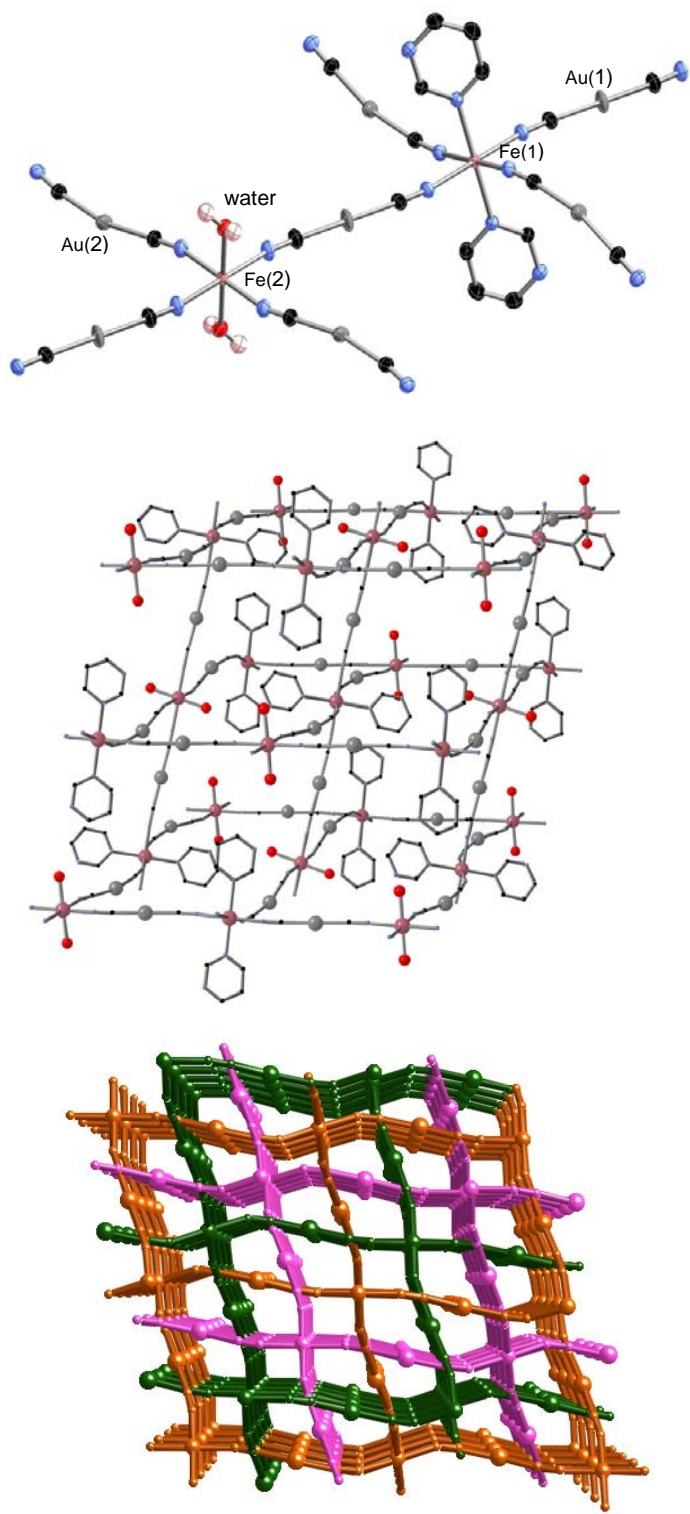

Fig. 1. Fragment of the coordination polymer [Fe(pmd)(H$_2$O){Au(CN)$_2$}$_2$]·H$_2$O displaying the asymmetric unit (top), the expanded version of the archetypal network structure of CdSO$_4$ (middle) and the triple interpenetration of identical networks (bottom). Color code: O, red; N, blue; C black; Au, grey; Fe, pink.

## Experimental details

The studied compound has been synthesized and characterized according to Ref. 23. An Ar/Kr ion laser was used for Raman excitation at 514.5 nm (2.41 eV) and 647 nm (1.92 eV). The laser output power was kept below 0.5 mW on a focus of approximately 100μm diameter to protect the sample from possible heating effects. All previous Raman experiments on spin-crossover systems that are known to us have been performed with at least 10 times large laser intensities. The scattered light was collected in quasi-backscattering configuration and dispersed by a triple monochromator DILOR XY on a liquid-nitrogen-cooled CCD detector. Measurements were performed in the parallel polarization configuration. Temperature dependencies were measured in a continuous helium flow cryostat from 10 K to room temperature. Experiments at elevated temperatures up to 370 K on the **2Au** dehydrated modification were carried out using a helium gas filled heating stage. The variable temperature magnetic susceptibility measurements were carried out in SQUID apparatus operating at 1 T and at temperatures 16 – 200 K. An externally placed frequency doubled Nd:YAG-laser ($\lambda$ = 532.1 nm) and optical fiber were used for light-irradiation under magnetic measurements.

The Raman and IR spectra were calculated with use of GAUSSIAN03 programme system [29] for HS and LS isomers of a trinuclear model of (**1Au**) with $C_i$ symmetry described later. Becke's three-parameter hybrid functional B3LYP, [30] which contains Becke's exchange functional [31] together with the local spin density correlation functional III of Vosko et al. [32] and the non-local correlation functional LYP of Lee et al. [33,34] was used for geometry optimization. As a basis set CEP-31G was used. [35-37]

## Results and discussion

For the monoclinic (space group P2$_1$/c, Z = 4) structure, the factor group analysis yields 156 Raman-active modes ($78A_g + 78B_g$) for the zone center vibrations. Due to the large number of allowed vibrations and the transparency of the irregular shaped samples a symmetry analysis of the observed modes in our Raman experiment has been omitted.

The observed excitation (Fig. 2) can be divided into three groups of Raman frequencies. Following earlier assignments[38-42] in other compounds with compositional similarity, these modes can be attributed mostly to internal modes of pyrimidine (600 - 1600 cm$^{-1}$), internal and external modes of {Fe(1)N$_6$} and {Fe(2)N$_4$O$_2$} cores (below 600 cm$^{-1}$), and C-N stretching-like vibrations (above 2150 cm$^{-1}$). This assignment is confirmed by the normal co-ordinate analysis performed within the framework of DFT, described further below.

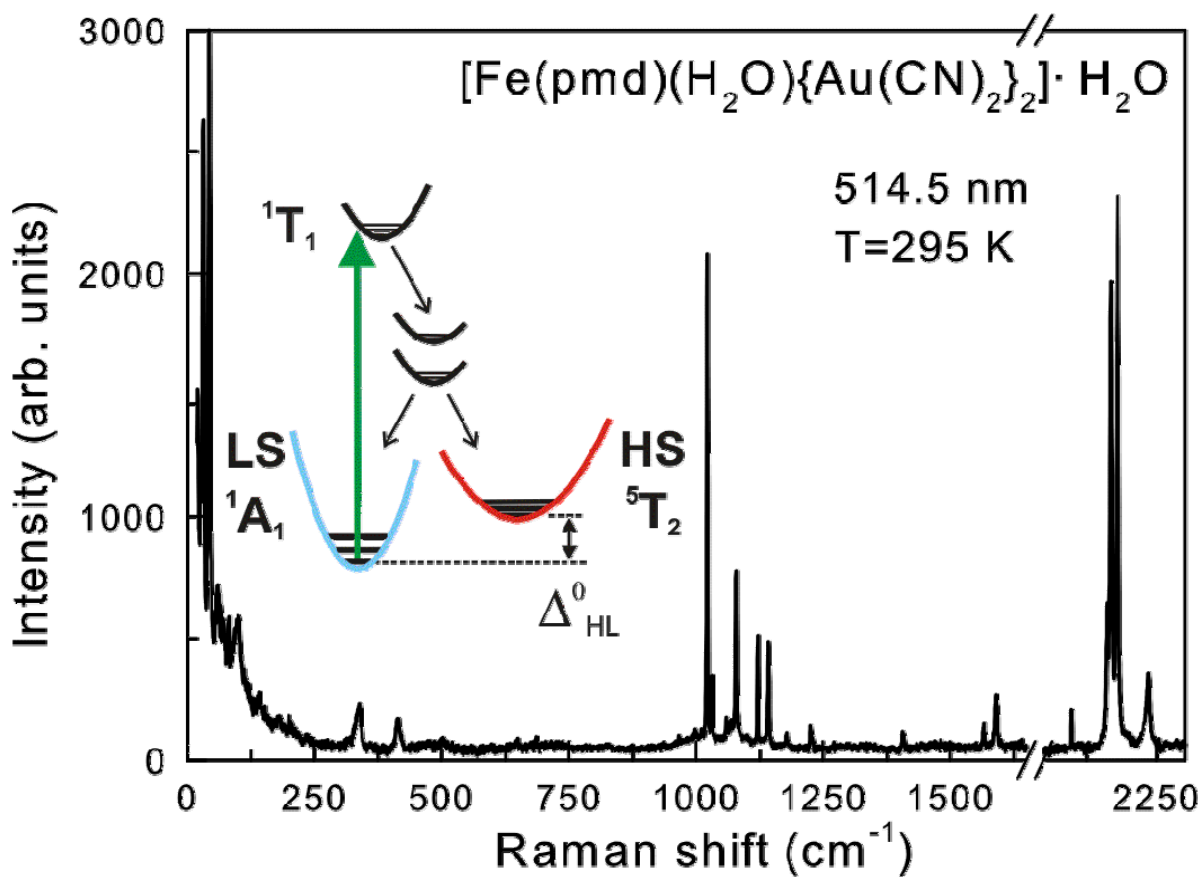

Fig. 2. Room-temperature Raman spectrum of a [Fe(pmd)(H$_2$O){Au(CN)$_2$}$_2$]·H$_2$O (**1Au**) single crystal. The inset shows a sketch of the spin state diagram including a light induced excitation with a 514nm laser line.

As we do not expect changes in the frequency region of the pyrimidine vibration, we will focus our attention to the high-frequency and to the low-frequency regions where remarkable changes are observed as function of temperature. We note that the relative change of the high-frequency modes with spin state is expected to be markedly smaller and that they have a smaller contribution to the evolution of entropy[43]. Representative Raman spectra of [Fe(pmd)(H$_2$O){Au(CN)$_2$}$_2$]·H$_2$O at different temperatures are presented in Figs. 3 and 4. A summary of the Raman spectral changes vs. temperature is presented in Table 1. It is clearly seen from Fig. 3 that a three-step transition takes place with reducing temperatures and that the effects involve both frequency as well as intensity of the modes in the low to intermediate frequency regime. Therefore these effects cannot be attributed to a change of optical constants, e.g. the optical penetration depth.

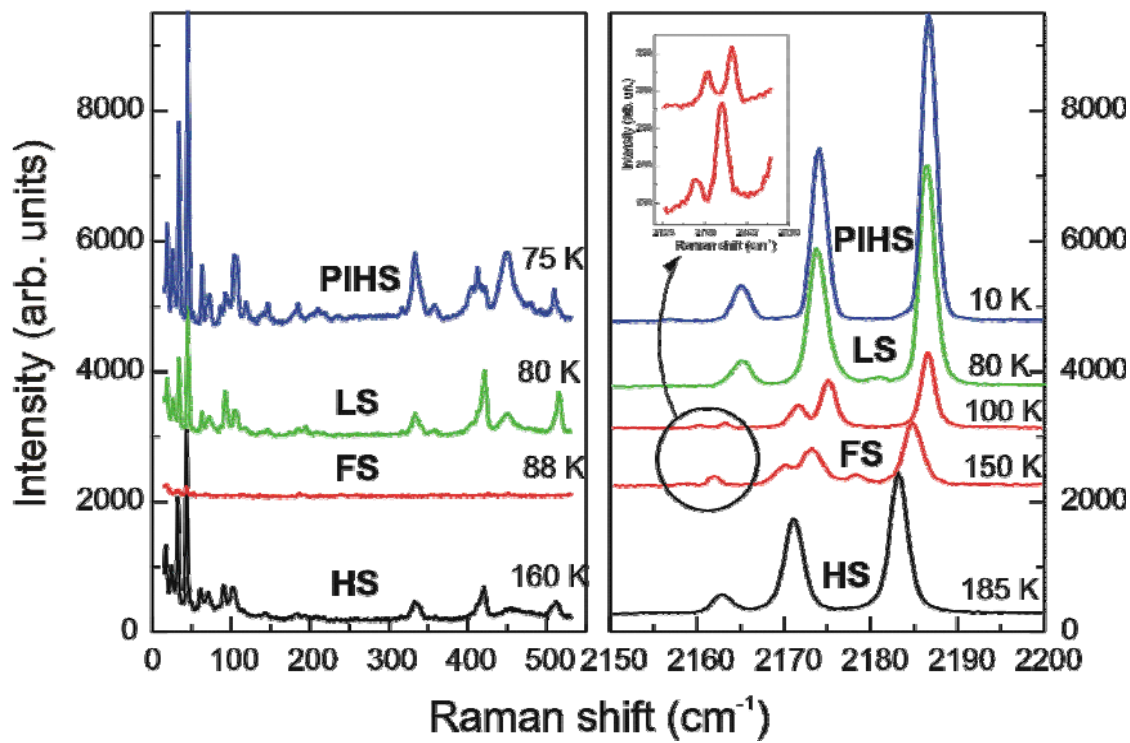

Fig. 3. Temperature dependent Raman spectra of the [Fe(pmd)(H$_2$O){Au(CN)$_2$}$_2$]·H$_2$O measured in the frequency regime of 0 – 550 cm$^{-1}$ and 2150 – 2200 cm$^{-1}$. The inset shows details of the spectra in the FS state phase.

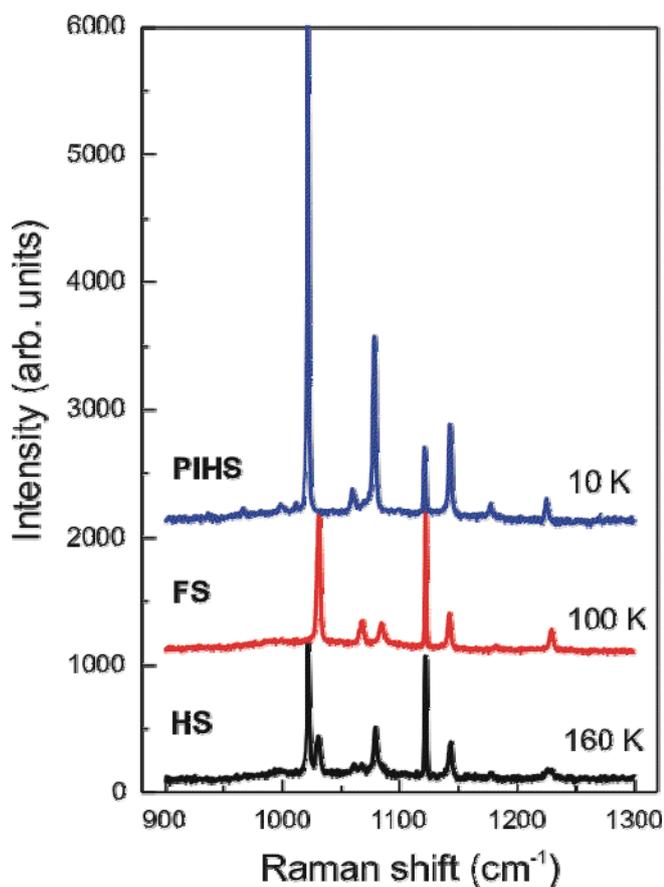

Fig. 4. Temperature dependent Raman spectra in the frequency region of internal modes of the pyrimidine.

| Temperature regimes (K) and spin state phase | Frequency regimes | | |
|---|---|---|---|
| | 10-600cm$^{-1}$ | 800-1700cm$^{-1}$ | 2150 cm$^{-1}$ and higher |
| | Internal and external modes of {Fe(1)N$_6$} and {Fe(2)N$_4$O$_2$} | Internal modes of Pyrimidine | C-N stretching like vibrations |
| 370 < T < 300 K | - | - | Structural change at ~345 K |
| 300 < T < 160 K **high spin (HS)** | No significant change in position | No significant change in position | All bands shift monotonically to high frequency |
| 160 < T < 88 K **fluctuating spin phase (FS)** | Abrupt decrease in intensity with 514.5 nm, not with 647 nm laser excitation | No significant change in intensity with 514.5 nm and 647 nm lasers | Few new modes arises and band position shifts |
| 88 < T < 80 K **low spin (LS)** | Intensity and line number recover, similar to HS | Intensity pattern similar to HS | Intensity pattern similar to HS |
| T < 80 K **photo-induced HS (PIHS)** | New bands appear | Intensity increases | No change |

Table 1. Summary of Raman spectral changes for different frequency regimes in the Fe$^{2+}$ coordination polymer [Fe(pmd)(H$_2$O){Au(CN)$_2$}$_2$]·H$_2$O as function of temperature and laser excitation. The structural change at ~345 K is related to the dehydration of the compound from (**1Au**) to (**2Au**) [23].

The first transition happens at temperature below 160 K. This is the critical temperature, $T_c$, where one of the two Fe$^{2+}$ sites in (**1Au**) undergoes a thermally induced first-order spin-crossover transition from the HS- to the LS-state [23]. In the following we will denote the corresponding phases according to the spin state of this Fe(1) ion site. Given that the HS and the LS phases of (**1Au**) have the same crystal symmetry, it is reasonable to expect no significant differences in their vibrational spectra. Instead just below $T_c$ we observe dramatic changes in the form of an abrupt loss of the Raman intensity in the low-frequency part of the spectra. Due to the magnitude of this intensity reduction it is even difficult to extract their intensity from the background noise.

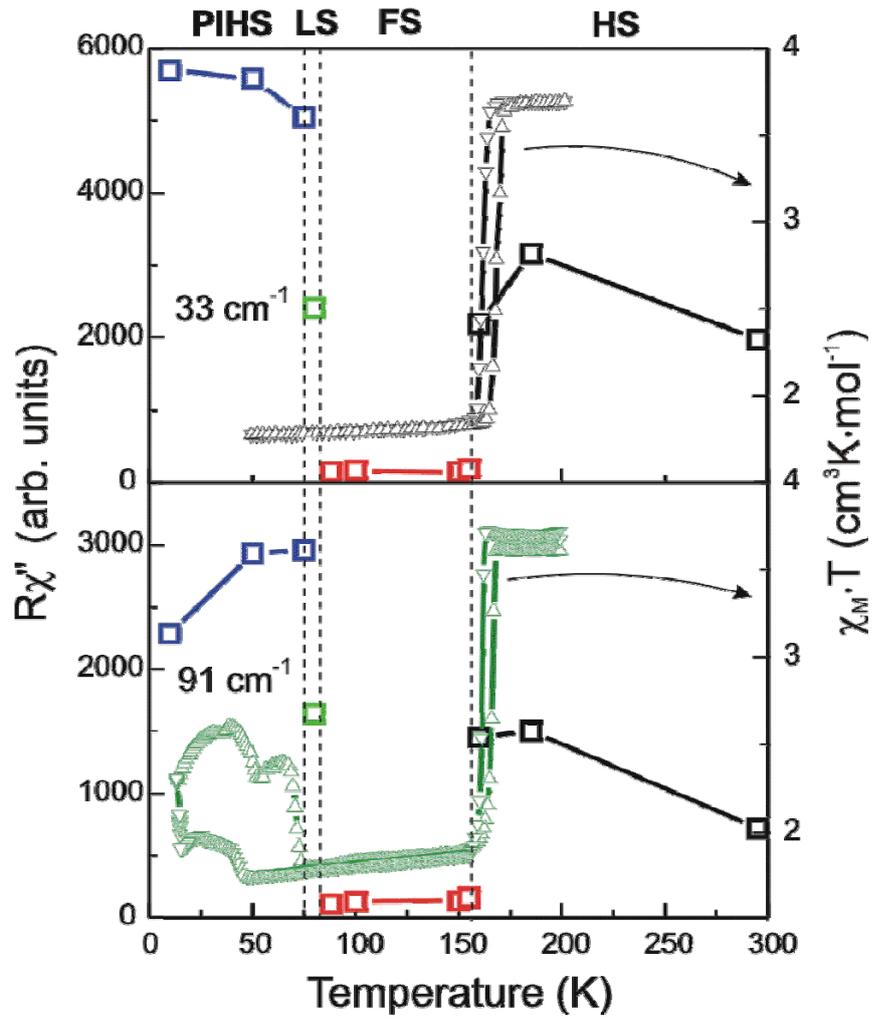

Fig. 5. Temperature dependence of integrated Raman intensities $R\chi''$ for selected low frequency phonon lines (33 and 91 cm$^{-1}$, open squares in upper and lower plot, respectively) and the susceptibility $\chi_M \cdot T$ without light irradiation [23] (triangles, upper right scale) and under laser ($\lambda$ = 540 nm) irradiation (triangles, lower right scale).

Figure 5 shows the behavior of the integrated intensity of selected low-frequency lines versus temperature. Usually, Raman spectra in the HS and LS states have approximately the same intensity[14,25,43] or the intensity of the LS-spectra is even higher [41]. In our case the intensity of low-frequency part of the spectra in the new phase is less than 5% from the intensity in of the HS-phase. Besides that, we observe changes in the number of lines as well as in their frequency (see Fig. 6). We note that the modes in the middle-frequency region (800 - 1700 cm$^{-1}$) have comparable intensities in both phases, while in the high frequency Raman signal the new phase shows up as two times smaller scattering intensity as in the HS phase. These spectral changes were found reversible and reproducible over several measurement cycles.

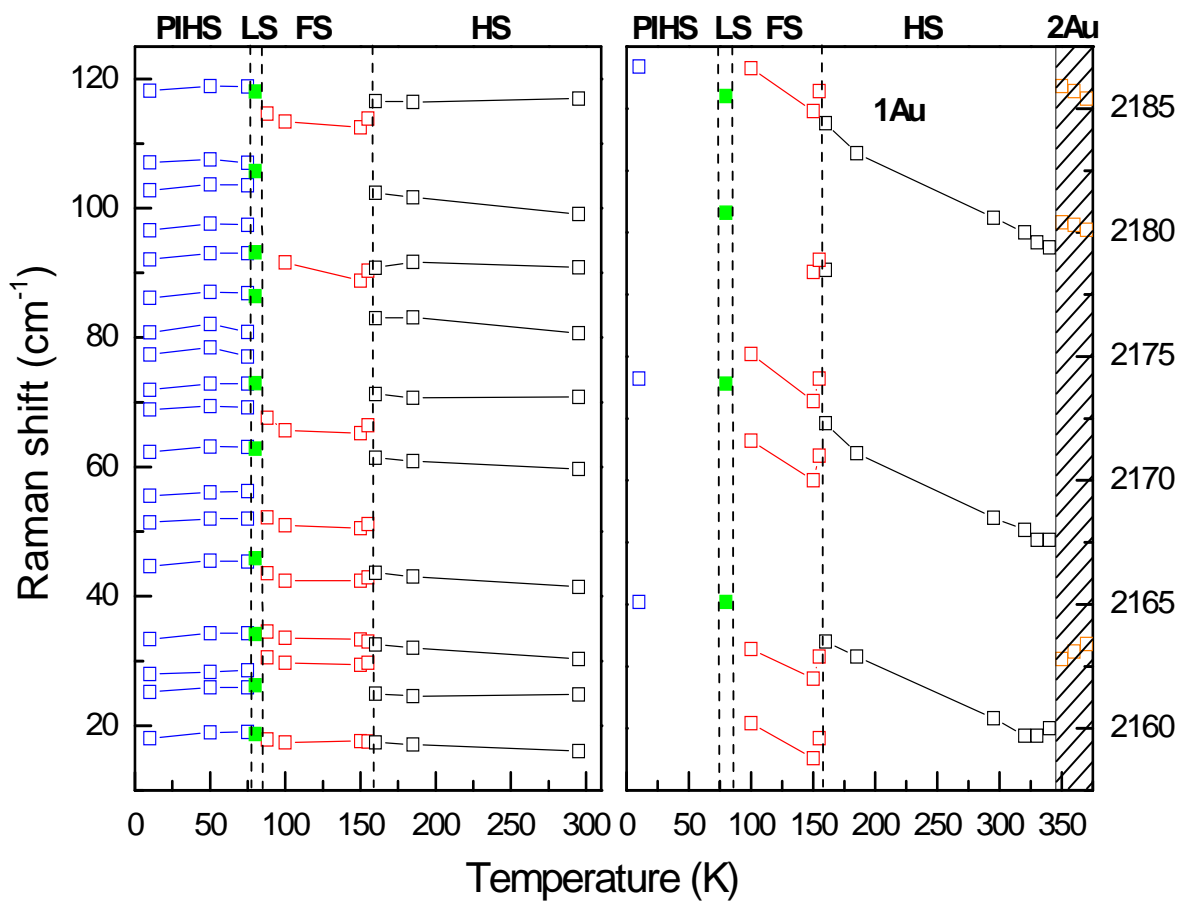

Fig. 6. Temperature dependence of the phonon line positions in the frequency regions of the {Fe(1)N$_6$} and {Fe(2)N$_4$O$_2$} octahedrons (left) and C-N stretching-like vibrations (right).

How to explain this unexpected result, namely a "wiping out" and sudden decrease of the RS intensity in the low frequency region? Checks for reproducibility with 514.5 nm excitation wavelength on different single crystals and the simultaneous investigation of low and intermediate frequency regime convinced us in its intrinsic nature. A review of recent results on spin-crossover systems shows that photo excitation and relaxation [44,45] processes are evident for some systems and complexities may develop if a competition between thermo- and light-driven processes exists on a molecular scale. The last aspect is most relevant in the temperature range where the zero-point energy difference between the LS and HS energy surfaces, $\Delta^0_{HL}$, is small (see inset in Fig.2). Dynamical spin state switching can be then be induced by any type of lattice distortions[47] as well as light irradiation. The resulting spin state – phonon interaction is strong and could even have non-adiabatic contributions as the time scale of interconversion ($\tau_{inter} \approx 80$fs) [46] of the excited singlet to the quintet state in Fe(II) complexes matches roughly with the period of typical phonons of the respective coordinations (400 cm$^{-1}$). As the evidence for a pure and static LS-state of the Fe(1) ions at temperatures below $T_c$ is based only on magnetization data without light irradiation [23] we will in the following assume a dynamic competition between temperature- and light-irradiation-driven processes in the Raman scattering experiments.

In the temperature range of 88 K < $T$ < 160 K we consider a dynamically fluctuating spin (FS) state with an incoherent switching between HS and LS states on different Fe(1) sites. The resulting incoherent variations of metal-ligand distances of the order of 0.1 Å are expected to lead to an anomalous damping of internal and external modes of {Fe(1)N$_6$} core. As to the Fe(2)N$_4$O$_2$ complexes, the influence of the surrounding medium, namely fluctuating internal pressure, can lead to a damping of phonon spectra, too. Just as to the spin state of Fe(2), the influence of the surrounding medium is not necessarily very dramatic [44]. Our aim should be to test this scenario by further experiments.

A further open issue concerns effects arising from a distortion of the octahedral symmetry. Most spin-crossover complexes have symmetry lower than octahedral and if the ligand field deviates strongly from octahedral symmetry even an intermediate spin state (triplet) can become the ground state [48]. A complex spin state transformation with the appearance of an intermediate spin-state of Co$^{3+}$ ions (3d$^6$-configuration) in a strongly distorted octahedral cage has been found recently in the rare-earth layered cobaltites RBaCo$_2$O$_{5.5}$ by muon relaxation experiment[49]. We do not think that a deviation from the octahedral symmetry due to the lattice strain is considerably strong for the Fe(2)N$_4$O$_2$ octahedrons but this as well as the effect of irradiation on Fe(2) spin state has to be clarified. Therefore, further structural investigations should be performed.

The mixed character of spin states displays itself in the high-frequency region of Raman spectra. It can be seen from Fig. 3 that going from the HS-state phase to the mixed fluctuating spin (FS) state phase leads to significant changes in the frequency region of C-N stretching-like vibrations. Besides the frequency shift, the lines at 2162 and 2172 cm$^{-1}$ split into two lines. The frequency behavior of high-frequency phonon lines is plotted in Fig. 6. A similar manifestation of mixed HS and LS states in the C-N stretching region has been observed earlier in an IR study in the spin-crossover system [50] [Fe(phen)$_2$(NCSe)$_2$].

We performed further Raman scattering experiments with a red-647nm laser line excitation. This energy is smaller than the $^1A_1 \rightarrow {}^1T_1/{}^1T$ transition. Therefore light-induced processes should be diminished. Figure 7 shows at three selected temperatures that the Raman scattering intensity does not drop through the thermally induced spin state transition.

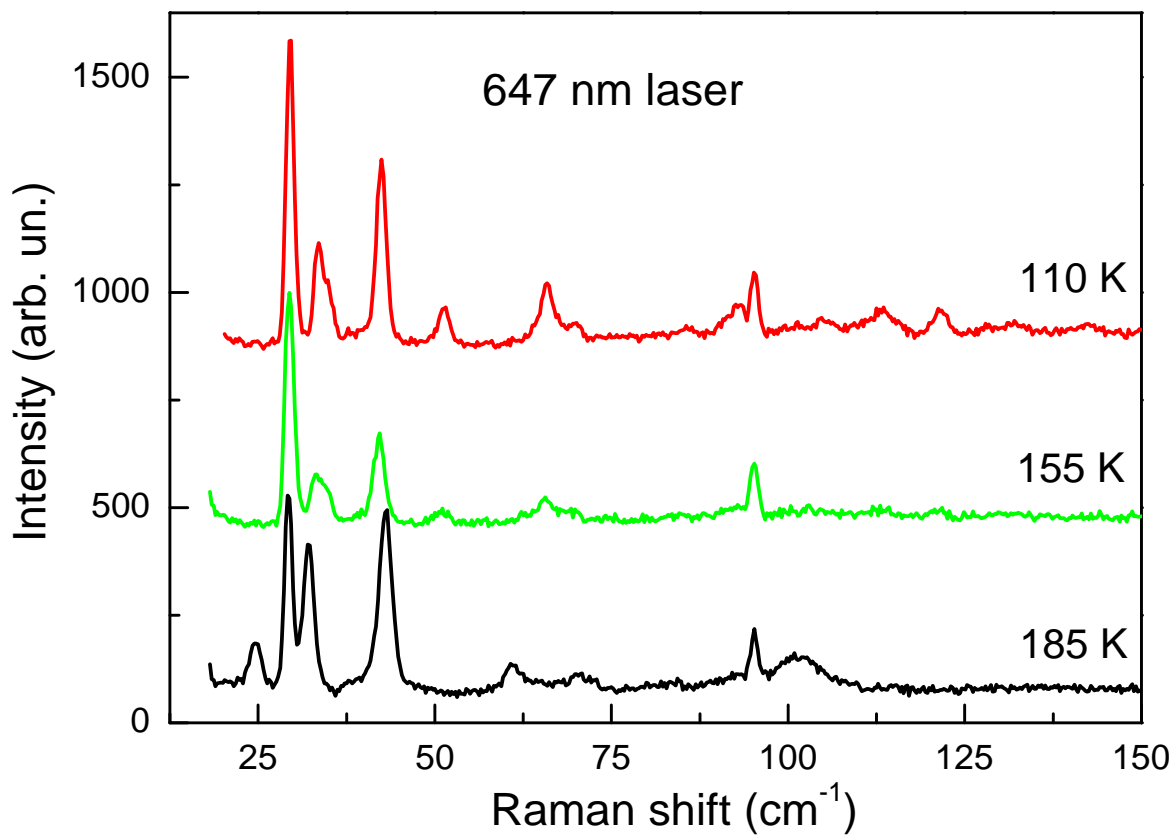

Fig. 7. Raman spectra of the [Fe(pmd)(H$_2$O){Au(CN)$_2$}$_2$]·H$_2$O excited by a red-647 nm laser line.

To our knowledge, this is the first observation of a "wiping out" effect in a spin-crossover system. Similar effects, anomalous damping and collapsing of phonon modes associated with dynamic phase fluctuation, have been observed earlier in La$_{0.7}$Ca$_{0.3}$MnO$_3$ [51]. However, our coordination polymer has a different, more complex spectrum of low frequency modes compared to the manganite. This could provide several competing relaxation paths leading to a glass like behavior. Analyzing recent efforts to investigate photo-excitation processes in spin-crossover compounds, we want to underline that all experiments reveal a complex picture of light irradiation action inside a thermal spin conversion, even at low temperatures [52]. Unfortunately, corresponding experiments at elevated temperatures on systems with a thermal hysteresis have not been fully successful. Still, unexpected and unexplained effects have been observed [45]. In this context, we highlight two interesting and unusual photoswitching phenomena recently observed in [Fe(pyrazine){Pt(CN)$_4$}], namely i) both the LS→HS and the HS→LS transitions were triggered by the same irradiation wavelength and ii) a mixed spin state was observed in a wide temperature range of the thermal hysteresis [53]. We also highlight that the excitation dependence and resonant nature of the "wiping out" effect in (**1Au**) is different from the resonant effects that have been observed in the spin-crossover system [Fe(pic)$_3$]Cl$_2$EtOH [54]. In the latter system an enhancement of only the ligand molecule phonon lines has been observed.

In the following we discuss the implications of the sequence of transitions and the related dynamics of the spin state for our understanding of the phonon wipeout: The second transition happens at temperature below 88 K. The Raman intensity increases again and the spectra become quite similar in number of lines, frequencies and intensities to those of the HS state. Therefore we assign this transition to a thermally induced LS-state. X-ray diffraction study did not reveal any crystal structure change in a wide temperature range [23], so the differences in the Raman spectra comparing the HS- and the LS-state are not expected and indeed are not observed, in the low frequency as well as in the high frequency region. The thermally-induced LS state is located in a very narrow temperature range of ~10 K.

The third and last transition occurs in the temperature region between 80 K and 75 K. While the high-frequency Raman spectra do not show (except small increases of frequency) any transformation under this spin state transition, a drastic change is observed in the low-frequency regime. A number of additional lines appear below 600 cm$^{-1}$ which are not observed at higher temperatures. Experiments on several samples with varying laser spot position and intensity show that this is an intrinsic and reproducible process. We attribute this transition to the light-induced $^1A_1$ (LS) → $^5T_2$ (HS) transition due to the LIESST effect. Our assignment is confirmed by SQUID measurements – a light-induced thermal hysteresis loop is observed at temperatures below 75 K (see Fig. 4, lower panel). The spectral change strongly indicates that a symmetry lowering takes place in the photo-induced phase. According to the model of *Tayagaki et al.* [14], the formation of the photo-induced HS-state occurs in the initial LS-state phase. Irradiation leads to the population of the $^5T_2$ (HS) metastable state (threefold orbital degenerate) via one or two intersystem crossing steps. The photo-induced phase is stabilized by a cooperative Jahn-Teller transformation [14], i.e. the symmetry lowering. As a result, formerly only IR-active modes appear in the Raman spectra of the low-temperature high-spin photo-induced phase.

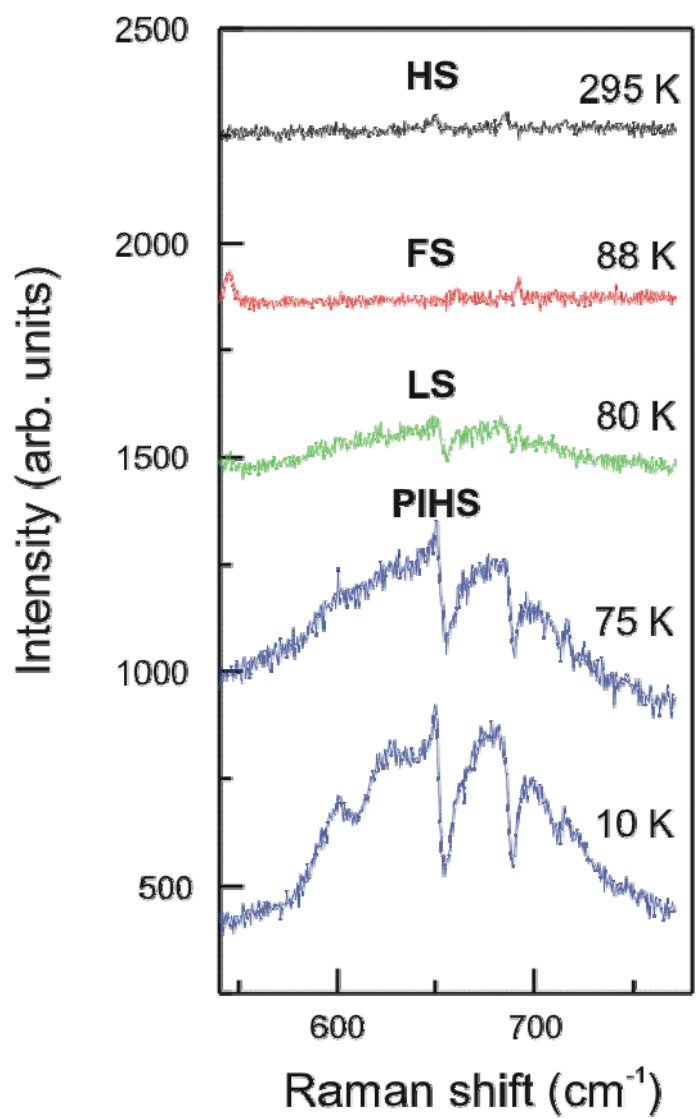

Fig. 8. Fano resonance in the PIHS phase of [Fe(pmd)(H$_2$O){Au(CN)$_2$}$_2$]·H$_2$O.

The presence of strong electron-phonon interaction manifests itself also directly in the Raman spectra (Fig. 8). A broad structured band centered at ~650 cm$^{-1}$ probably due to scattering on electronic excitations appears in the low-temperature spectra. Electron-phonon coupling leads to a very strong Fano resonance at 650 and 690 cm$^{-1}$, frequencies attributed to Jahn-Teller-like and breathing-like phonon modes of FeN$_6$ octahedra, respectively, with pyrimidine in-phase bending coupled to Fe-N stretching. To our knowledge, this it is the first observation of a Fano resonance in the light-induced phase of spin-crossover systems.

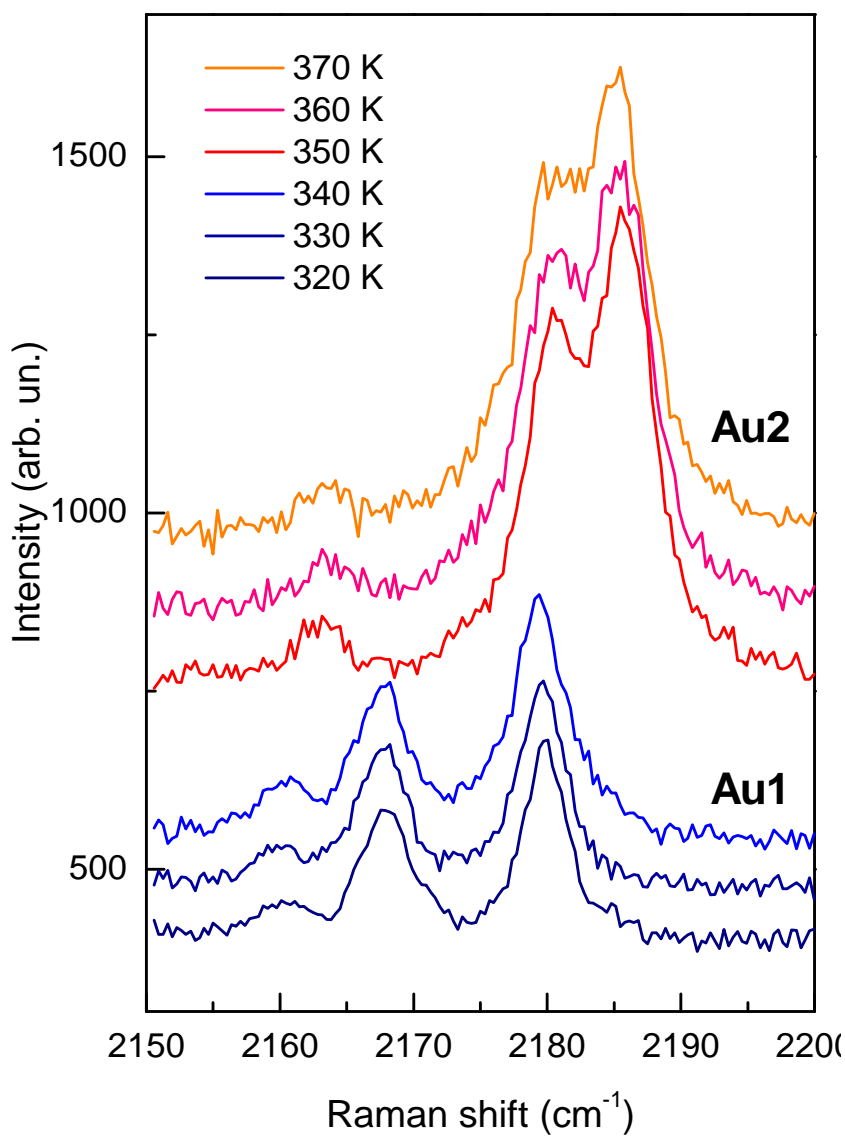

Fig. 9. High-temperature Raman spectra of [Fe(pmd)(H$_2$O){Au(CN)$_2$}$_2$]·H$_2$O single crystal in the frequency region of C-N stretching-like vibrations.

Finally we have studied Raman spectra under heating the coordination polymer. In the temperature range 323-382 K the set of interpenetrating and independent 3D networks in (**1Au**) is transformed to the single 3D network (**2Au**) by losing the coordinated and nonbonded water molecules[23]. This transformation leads to significant structural changes: the (pmd) ligand now bridges directly the Fe(1) and Fe(2) sites defining a system of infinite chains {-Fe(1)-pmd-Fe(2)-} running parallel to the *a* axis. The [Au(CN)$_2$] groups of one chain link with the equatorial positions of the iron centers, connecting adjacent chains and defining a single 3D network. Raman spectra at selected temperatures in the frequency region of 2150 – 2200 cm$^{-1}$ are presented in Fig. 9. The measurements were performed after first heating the sample to T = 370 K and then cooling it down to the given temperatures. Various measurements were done during heating and cooling of the sample to proof reproducibility of the dehydration effect. Within the described temperature protocol also no hysteresis was found. The structural changes at ~345 K lead to a jump-like shift of CN stretching modes at 2160 and 2179 $cm^{-1}$ to the positions of 2163 and 2180 $cm^{-1}$, while the 2167 $cm^{-1}$ mode shifts quite a lot to the position of 2186 $cm^{-1}$. Such different behavior point to the nonequivalent positions of CN groups in the crystal structure.

## DFT Calculations of a spin state coordination

The DFT calculations have been performed in order to allow a decisive assignment of the relevant observed Raman bands. The DFT approach was recently shown to be quite an effective method to predict the vibrational properties of spin-crossover systems [55-58]. For the system under study, because of its polymeric nature, the calculations shall in principle involve the complete cell under periodic boundary conditions. However with the available software allowing for periodic boundary conditions, the calculations for equally large and complex systems are hardly feasible. Therefore we decided to use the Gaussian programme that offers high-end functionals and basis sets for a model system, which is as relevant as possibly to the **(1Au)** complex. As we are mainly interested in vibrations that are dependent on the spin state transition, we chosed a model that reproduces the environment on the spin-crossover Fe(II) (Fe(1)) in the best possible way. On the other hand, in order to keep the calculations feasible we had to limit the size of the molecule. Therefore the calculations have been performed for a trinuclear unit of $C_i$ symmetry encompassing a Fe(pmd)$_2$([Au(CN)$_2$]$_4$ fragment with two Au(CN)$_2$ bridging to the two terminal Fe([Au(CN)$_2$]$_4$(H$_2$O)$_2$. Therefore an anionic unit with stoichiometry Fe$_3$[Au(CN)$_2$]$_{10}$(H$_2$O)$_4$(C$_4$H$_4$N$_2$)$_2$ and a charge of -4 was used for calculations. The system looks like the fragment shown in Fig. 1 (top) with one more FeN$_4$O$_2$ fragment centered on Fe(1) attached, so that the Fe(2) atom lies in the symmetry center.

An initial test of the quality of the calculation results may be the comparison of the obtained metal-ligand bond lengths for the spin-crossover iron centre. These values are listed in the Table 2.

| Calculated (DFT) bond lengths (Å) | | Bond lengths from X-ray data for (**1Au**) (Å) |
|---|---|---|
| Fe-N(pyrimidine) | 2.215 HS | 2.200 HS |
| | 2.024 LS | 1.986 LS |
| Fe-N (NC – bridg.) | 2.182 HS | 2.149/2.164 HS |
| | 1.970 LS | 1.930/1.948 LS |
| Fe-N (CN-nonbridg.) | 2.142 HS | ------ |
| | 1.967 LS | |
| Au-C | 2.016 HS | 1.992/1.989 HS |
| | 2.016 LS | 1.976/1.981 LS |

Table 2. Calculated (model coordination) and experimental metal-ligand bond-lenghts[23] around the spin-crossover iron centre of the model complex.

The above data reveal that the calculated *in vacuo* values show a general bias towards longer bond lengths compared to the solid state data. However the observed differences are less than 0.04Å and therefore reasonable vibrational frequencies may be obtained in the following normal co-ordinate analysis.

The normal coordinate analysis yields 249 normal vibrations that are of $A_g$ (Raman active) and $A_u$ (IR active) type. Those relevant to the above presented spectra are listed in table below.

| Mode type | LS (cm$^{-1}$) | HS (cm$^{-1}$) |
|---|---|---|
| Au-C-N bending | 305-360 | 306-355 |
| Fe-N-Au-C (no CN stretching) stretching around Fe(1) and Fe(2) (with Fe(1)-N$_4$ breathing) | 523 | 489 |
| Fe(2)-N –Au-C (no CN stretching) stretching | 492 | 499 |
| pyrimidine in-phase bending + Fe(1)-N(pyrimidine) stretching | 644 | 632 |
| pyrimidine in-phase bending + Fe(1)-N(pyrimidine) stretching | 690 | 678 |
| pyrimidine ring bending + Fe-N stretching | 997 | 985 |
| pyrimidine ring bending | 1066 | 1065 |
| pyrimidine C-H in-plane bending | 1085 | 1074 |
| pyrimidine C-H in-plane bending | 1144 | 1141 |
| pyrimidine C-H in-plane bending | 1221 | 1215 |
| pyrinidine C-H in-plane bending | 1237 | 1230 |

| pyrimidine ring stretching | 1603 | 1593 |
| N-C-Au stretching around Fe(2) | 2179 | 2168 |
| Fe(1)-N-C-Au(axial) (CN stretching) stretching | 2185 | 2174 |
| Fe(1)-N-C-Au(axial) (CN stretching) stretching | 2196 | 2183 |

Table 3. The calculated Raman active $A_g$ vibrations, relevant to the observed Raman spectra.

Generally, the calculations reveal a pattern that is in line with the observed spectra. On the other hand they shed light on the problem why the changes in the observed spectral pattern on LS-HS transition are so subtle. It's generally accepted [59] that a change of a spin state in Fe(II) nitrogen complexes results in a shift of metal-ligand stretching vibrations by 100-200 cm$^{-1}$. Yet in our case no Raman shift is observed in the 200-500 cm$^{-1}$ regime and the calculated Fe-N(CN) vibrations show a limited shift of a few tenths of a wavenumber. This may be due to the fact that every Fe-N(CN) stretching is also a Au-C (stretching) and therefore its frequency is primarily dependent on a more covalent component. On the other hand, although the calculated Fe-N(pyrimidine) stretching modes show frequency shifts on the LS-HS transition (HS $A_u$: 186, 204, 242; LS $A_u$: 402, 492 ) the modes are of ungerade symmetry and are therefore not observable in Raman spectra. They might become Raman active if the centre of symmetry of the molecule is cancelled. For the polymeric system (**1Au**) this may happen when two different spin states of Fe(1) are present, i.e. if the cooperative Jahn-Teller transformation occurs.

## Conclusion

Using Raman spectroscopy we have demonstrated that the spin state conversion of the Fe(1) site in the cyanide-based bimetallic coordination polymer [Fe(pmd)(H$_2$O){Au(CN)$_2$}$_2$]·H$_2$O follows a complex sequence of high spin, fluctuating spin, low spin and photo-induced high spin states. A novel "wiping out" effect was observed in the temperature range of 88 K < $T$ < 160 K. We attribute this effect to a competition of thermo- and light-driven processes in the temperature range with small $\Delta^0_{HL}$ and a matching of the interconversion time scale with the period of the involved vibrations. This competition leads to the creation of a complex spin-state fluctuating phase and, as a result, to the anomalous damping of internal and external modes of the {Fe(1)N$_6$} complexes. Our experimental and theoretical data are fully consistent with an intrinsic origin of the fluctuating spin state. This is in contrast with recent findings in a Prussian Blue analog with a stoichiometry distribution[60]. At low temperatures the PIHS phase in the coordination polymer appears due to the

LIESST effect. The additional lines which are neither observed in the HS nor in the LS phases indicate that the vibrational selection rules are modified in the photo-induced phase.

**Acknowledgments:** We acknowledge important discussions with P. Gütlich, J. J. McGarvey, and G. G. Levchenko. This work has been supported by the DFG Priority Program SPP1137 on *Molecular Magnetism* and the network of the European Science Foundation *Highly Frustrated Magnetism*. V. G. and Yu. G. P. acknowledge the support of the Ukrainian-Russian grant 2008-8. V.S. and J.A.W. acknowledge the support of the BMBF under 05KS7UK2